\documentclass[12pt]{iopart} 

\usepackage{graphicx}
\usepackage{amssymb} 

\begin{document}
\title{Structure, stability and defects of 
single layer \textit{h}-BN in comparison to graphene}
\author{GJ Slotman and A Fasolino}
\address{Institute for Molecules and Materials, Radboud University Nijmegen, Heyendaalseweg 135, 6525AJ Nijmegen, The Netherlands}
\ead{a.fasolino@science.ru.nl}

\begin{abstract}
We study by molecular dynamics the structural properties of single layer \textit{h}-BN in comparison to graphene. We show that the Tersoff bond order potential developed for BN by Albe, M\"oller and Heinig gives a thermally stable hexagonal single layer with a bending constant $\kappa=0.54$ eV at $T=0$.  We find that the non-monotonic behaviour of the lattice parameter, the expansion of the interatomic distance and the growth of the bending rigidity with temperature are qualitatively similar to those of graphene. Conversely, the energetics of point defects is extremely different: instead of Stone Wales defects, the two lowest energy defects in \textit{h}-BN involve either a broken bond or an out of plane displacement of a N atom to form a tetrahedron with three B atoms in the plane. We provide the formation energies and an estimate  of the energy barriers. 
\end{abstract}

\section{Introduction}
The interest in two-dimensional (2D) crystals is rapidly extending to other materials than graphene, often combined to form man-made heterostructures\cite{britnell2012}. Hexagonal boron-nitride (\textit{h}-BN), which has similar structural properties to graphene but is an insulator with a gap of $\sim5-6$ eV\cite{Goldberg2010}, is one of the promising dielectric materials for integration in hybrid graphene devices\cite{Bresnehan2012}. Deposition of graphene on \textit{h}-BN is found to improve the transport properties of graphene possibly due to suppression of scattering from out of plane ripples\cite{Dean2010}. The growth of thin BN films is known to require a certain amount of ion irradiation which in turn leads to the detrimental formation of point defects\cite{Peter2009}. The presence of defects lead to special features in x-ray core level spectroscopy\cite{Peter2009, Pavlychev1998} which have stimulated the study of selected defects in \textit{h}-BN by first-principles\cite{ma2009stone,chen2009electronic,wang2012,okada2009}. Another interest in the defect structure of \textit{h}-BN is the possibility to create suitable chemisorption sites for molecules as proposed for graphene\cite{sanyal2009molecular}.  In this paper, we address the structural stability, thermal expansion and defect formation in \textit{h}-BN by means of Molecular Dynamics (MD) based on a classical description of interatomic interactions. The success of this approach  for graphene, particularly for temperature dependent properties, has been due to the existence of several accurate models of interactions\cite{tersoff1988empirical,
brenner2002second,stuart2000reactive,los2005improved}. These so called bond-order potentials for carbon, pioneered by Tersoff\cite{tersoff1988empirical}, have the important feature of being reactive, namely to allow change of coordination.  For BN, Albe, M\"oller and Heinig\cite{albe:1997} have developed a Tersoff potential that is supposed to describe well the bulk BN phases and the defect formation energy and has been used to study the effect of irradiation in single layer \textit{h}-BN\cite{lehtinen2011}. To our knowledge, however, no comprehensive study of the structural stability and defect formation energies based on this approach exists up to date.  The goal of this article is therefore two-fold: on the one hand, to study the structure of single layer \textit{h}-BN in comparison to graphene and report an extensive search for low energy point defects and, on the other hand, provide a wide set of results that can be later tested against experiments or more sophisticated calculations. Validation of this potential is necessary also in view of a possible extension to a reliable potential for hybrid systems formed by B, N and carbon. We find that the thermal expansion and bending rigidity of \textit{h}-BN behave similarly to graphene whereas the type and formation energy of lattice defects is very different. The paper is organized as follows: in section~\ref{simulation} we describe our method and studied samples, in section \ref{a-par} and \ref{bendingkappa} we present our results for the temperature dependence of thermal expansion and bending rigidity of 
crystalline single layer \textit{h}-BN respectively. In section \ref{melting} we show how, by raising the temperature up to melting, we identify several possible lattice defects for which we give the formation energy. 
In section~\ref{conclusion} we give a summary and perspectives of our work.

\section{Method}
\label{simulation}
We have performed MD simulations using the \textsc{lammps}~\cite{lammps} software package and the BN interatomic potential of Albe, M\"oller and Heinig~\cite{albe:1997}. 
This potential gives at $T=0$ the  \textit{h}-BN equilibrium lattice parameter  $a=\sqrt3 R_0= $2.532 \r{A} \cite{albe:1997} where $R_0$ is the B-N interatomic distance. Experimentally, the lattice parameter of \textit{h}-BN is  
$2.504$ \r{A} and decreases with increasing temperature~\cite{paszkowicz2002lattice}. This means that the potential overestimates this value, resulting in a larger mismatch of $2.9$\% with the lattice parameter of 2.46 \r{A} for graphene, instead of the $1.8$\% experimental value.

To study the thermal stability and lattice expansion we have performed simulations starting with a flat sheet of \textit{h}-BN consisting of $n=9800$ atoms (sample A) corresponding to sample sides  $L_x=177.3$ \r{A} and $L_y=153.5$ \r{A} and periodic boundary conditions.  These simulations were performed in the isobaric-isothermal (nPT) ensemble using a Berendsen thermostat with damping parameter $\tau_t = 0.1$ ps to control the temperature and  a Berendsen barostat \cite{berendsen:3684} to impose an external pressure $P=0$. A smaller sample of $n=1296$ atoms ($L_x=45.59$ \r{A} and $L_y=78.96$ \r{A}) (sample B) was used to study defects by performing simulations in the canonical (nVT) ensemble using the $T=0$ lattice parameter. We have found that simulations with the Nos\'e-Hoover thermostat yielded similar values for the lattice expansion
but had undesired fluctuations in the enthalpy. The time step of all simulations was set to $\Delta t = 0.1$ fs and integration was performed using the standard velocity-Verlet algorithm. For each simulation with sample A, a total of 2.5 millions time steps were taken, where the last 2.0 millions ($200$ ps) were used for averaging.

\section{Lattice Parameter}
\label{a-par}

\begin{figure}[h!]
\begin{center}
\includegraphics[angle=0,width=0.5\textwidth]{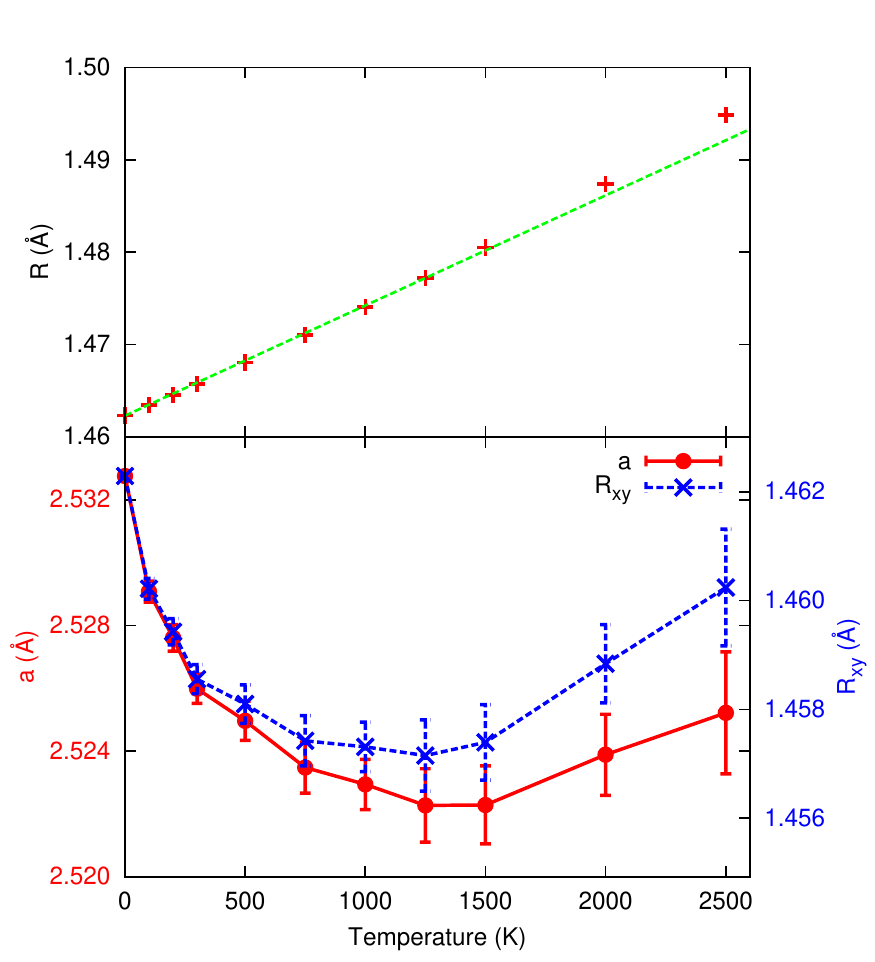}
\end{center}
\caption{Temperature dependence of lattice parameters calculated by MD in the nPT ensemble for sample A consisting of $n=9800$ atoms. (Top panel) Average BN interatomic distance $R$. A linear fit to the calculated points with $R=R_0+\alpha T$ where $R_0=$1.462 \AA~ and $\alpha=1.2\cdot10^{-5}$ \r{A} $K^{-1}$ describes well the results up to  $\sim$ 1500K. (Bottom panel) Lattice parameter $a$ (left y-axis) and BN interatomic distance projected in the xy-plane $R_{xy}$ (right y-axis). Notice that left and right y-axis differ by a factor $\sqrt{3}$.}
\label{fig:nearest_neigh}
\end{figure}

One of the interesting quantities is the lattice thermal expansion. For graphene, simulations  based on the bond order potential LCBOPII~\cite{PhysRevLett.102.046808} have found that the lattice parameter first decreases with increasing temperature up to $\sim $900 K and then increases at higher temperature. Calculations based on the quasi-harmonic approximation predict instead a contraction of the lattice parameter at least up to 2500 K\cite{MounetMarzariPRB}. Experimentally, data up to 400 K for graphene~\cite{Bao} and up to room temperature for \textit{h}-BN~\cite{paszkowicz2002lattice} confirm a contraction. We distinguish three different structural parameters, namely the in-plane lattice constant $a$ determined from the equilibrium box size obtained at constant pressure $P=0$, the B-N nearest neighbour distance $R$ and the nearest neighbour distance projected in the $xy$-plane $R_{xy}$.
At $T=0$ K,  for  a completely flat sheet of \textit{h}-BN,  $R=R_{xy}= a / \sqrt3$.
In \fref{fig:nearest_neigh}  we show the calculated temperature dependence of these three quantities for single layer \textit{h}-BN. We find that the BN nearest neighbor distance increases linearly up to about $1500$ K with a slope of $1.2\cdot10^{-5}$ \r{A} K$^{-1}$, which is almost twice as large as the value $6.5\cdot10^{-6}$ \r{A} K$^{-1}$ found in~\cite{PhysRevLett.106.135501} for freestanding graphene using \textit{ab initio} MD simulations.
For the lattice parameter we find thermal contraction up to $1200$ K, similarly to prediction for graphene\cite{PhysRevLett.102.046808}. For higher temperatures, the lattice parameter grows less rapidly than for graphene and, in the studied temperature range up to $2500$ K, it never gets larger than the value at $T=0$ K. At $T\ne 0$, the lattice parameter $a$ is not equal to $\sqrt{3}R_{xy}$ due to fluctuations in the out of plane direction.

\section{Bending rigidity $\kappa$}
\label{bendingkappa}

We analyse the height fluctuations  $h$ of the single layer \textit{h}-BN crystal by using the theory of membranes in the continuum (see for instance \cite{nelson2004statistical,katsnelson2012graphene,fasolino2007intrinsic,zakharchenko2010atomistic,katsnelson2012accounts}). Key quantities in this theory are the correlation functions of the height fluctuations and of the normals. The normal-normal correlation function $G(q)$ in the harmonic approximation is given by\cite{fasolino2007intrinsic,katsnelson2012accounts}:
\begin{equation}
G_0(q) = \left< \left| \mathbf{n_q} \right|^2 \right> = \frac{1}{S} \frac{k_B T n}{\kappa q^2}
\label{eq:gq}
\end{equation}
where $S = L_x L_y / n$ is the area per atom, $\kappa$ is the bending rigidity and the subscript zero indicates that averages are taken in the harmonic approximation, namely neglecting the coupling of in-plane and out of plane fluctuations in the stress tensor~\cite{katsnelson2012graphene}. In the limit of slowly varying height fluctuations, one can show that $G(q)$ is related to the height-height correlation function $H(q)$ by:
\begin{equation}
 G(q) = q^2 H(q)\equiv q^2\left< \left| h_{\mathbf{q}} \right|^2 \right>
\label{eq:gq_prop_hq}
\end{equation}
yielding, in the harmonic approximation,
\begin{equation}
\label{eq:hq}
 H_0(q) = \frac{1}{S} \frac{k_B T n}{\kappa q^4}
\end{equation}

\begin{figure}[h!]
\begin{center}
\includegraphics[angle=0 ,width=0.5\textwidth]{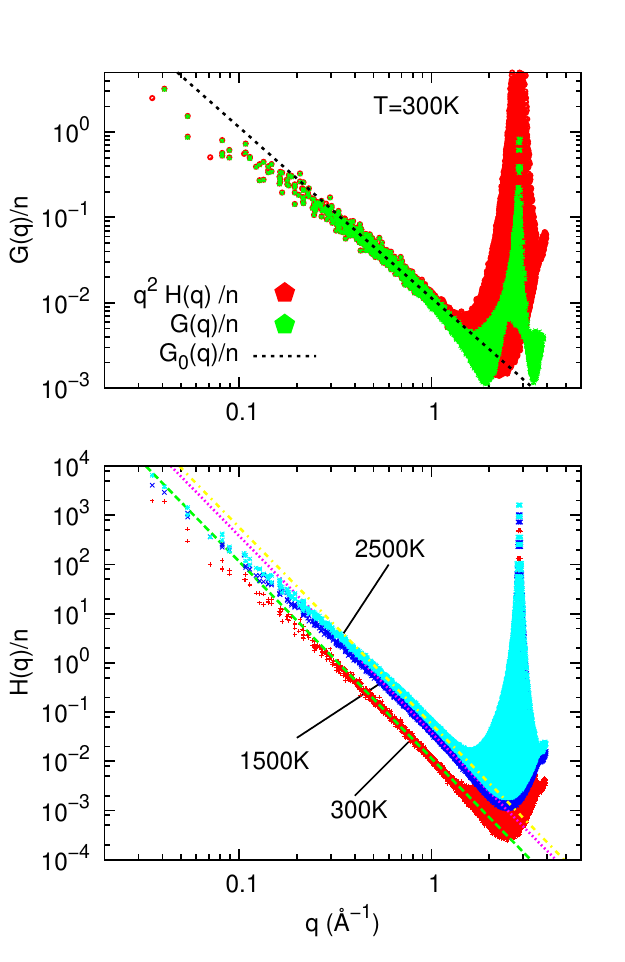}
\end{center}
\caption{(Top panel) The two correlation functions $G(q)/n$ \Eref{eq:gq} and $q^2 H(q)/n$ \Eref{eq:hq}  at $T=300$K coincide as expected from \Eref{eq:gq_prop_hq} for wave vectors  $q \lesssim  1$ \r{A}$^{-1}$ yielding the same value of $\kappa$ from a best fit to the harmonic approximation \Eref{eq:hq} (dotted line) in the range $q=0.5-0.9$ \r{A}$^{-1}$. Deviations from power-law behaviour occur for $q \gtrsim 1$ \r{A}$^{-1}$ close to the Bragg peak position at $q = 4 \pi / \sqrt 3 a = 2.86$ \r{A}$^{-1}$.  In the long wavelength limit, $q < q^*\approx 0.24$ \r{A}$^{-1}$, the correlation functions have a power-law behaviour with a smaller exponent\cite{los2009scaling}. (Bottom panel) $H(q)/n$ for $T=300$ K, $T=1500$ K and $T=2500$ K.}
\label{fig:gq_vs_hq_temp}
\end{figure}

It has been shown\cite{zakharchenko2010atomistic} that \Eref{eq:gq_prop_hq} is well reproduced when using numerical results from atomistic simulations only if the height fluctuations are calculated by averaging over the nearest neighbours heights:
\begin{equation}
h = \frac{1}{2} \left(h_0 + \frac{1}{3} (h_{\alpha} + h_{\beta} + h_{\gamma} ) \right)
\end{equation}
where $h_0$ is the z coordinate of one atom and $h_i$ the z coordinates of its three nearest neighbours. 
In the top panel of \fref{fig:gq_vs_hq_temp}, we compare $G(q)$ to $q^2H(q)$ calculated by MD at $T=300$ K.  We notice that these functions indeed coincide for $q\lesssim  1$ \r{A}$^{-1}$. Above this value the continuum approximation used in the theory of membranes breaks down and deviations from power-law behaviour occur, resulting in a peak at the position of the Bragg peak $q = 4 \pi / \sqrt 3 a = 2.86$ \r{A}$^{-1}$.  

The theory of membranes\cite{katsnelson2012graphene,katsnelson2012accounts} predicts deviations from harmonic behaviour for wavevectors smaller than
\begin{equation}
q^*= \sqrt{\frac{3TY}{8\pi\kappa^2}}
\end{equation}
where $Y$ is the bulk modulus. In the longwavelength limit, for $q<q^*$,  the correlation functions bend to a lower exponent. This feature reduces the divergence of out of plane fluctuations and stabilizes the 2D crystal\cite{katsnelson2012graphene,nelson1987fluctuations}. In the top panel of \fref{fig:gq_vs_hq_temp} we see 
that deviations of the calculated points from $G_0(q)$ start at $q^*\approx 0.24$ \r{A}$^{-1}$ as found for graphene at the same temperature\cite{katsnelson2012accounts,los2009scaling}, suggesting a similar ratio $Y/\kappa$.

\begin{figure}[h!]
\begin{center}
\includegraphics[angle=0,width=0.4\textwidth]{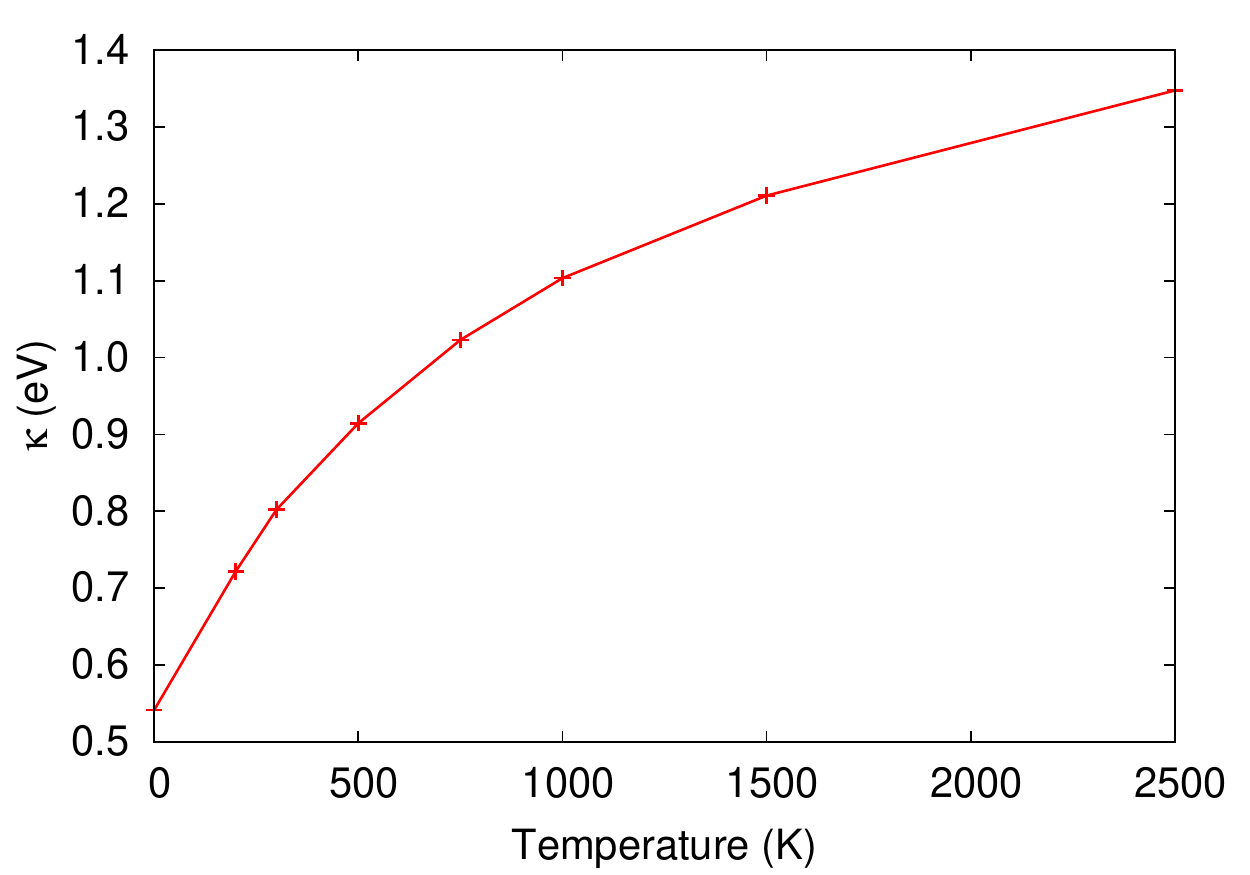}
\end{center}
\caption{Bending rigidity $\kappa$ as a function of temperature. The $T=0$ K value $\kappa=0.54$ eV is found by direct total energy calculations (see text).}
\label{fig:kappa}
\end{figure}

In the bottom panel of \fref{fig:gq_vs_hq_temp} we show $H(q)/n$ calculated by MD at different temperatures. 
The bending rigidity $\kappa$ can be obtained by a fit of the slope of the calculated $H(q)$ to \Eref{eq:hq} in the range of $q$ vectors where the harmonic approximation applies, yielding the temperature behaviour shown in \fref{fig:kappa}. The value of $\kappa$ at $T=0$ can be found by direct total energy calculations of nanotubes and extrapolation to the limit of infinite radii\cite{tersoff1994structural}. This procedure yields $\kappa=0.54$ eV, a value lower than the value $\kappa = 0.82$ eV for graphene\cite{fasolino2007intrinsic}, meaning that \textit{h}-BN is easier to bend. 

\section{Defects}
\label{melting}

Classical MD simulations are very suitable to search for possible distortions of the lattice. By raising the temperature  we  have  observed the formation of some defects that arise in the melting process which occurs spontaneously at  $T\sim 4000$ K. We have then studied the energetics of these defects and of others defects that are known to occur in graphene and in semiconductors. We distinguish two kinds of point defects,  those where the number of atoms remains the same and those where one or more atoms are removed (vacancies). For the first group of  defects we can calculate the formation energy as:
\begin{equation}
\label{eq:for_energy}
 E_F = E_{defect} - E_{perfect} ,
\end{equation}
where $E_{defect}$ and $E_{perfect}$ are the total energies of the sample with and without the defect. The cohesive energy is $E_{coh}\equiv E_{perfect}/n=-6.4166$ eV.  For these calculations we use a sample of $n=1296$ atoms (sample B) which is cooled down to $T=0$ K in the nVT ensemble. In these calculations all defects were artificially created, thus not always observed in high temperature simulations.

\begin{figure}[h!]
\begin{center}
\includegraphics[angle=0,width=0.35\textwidth]{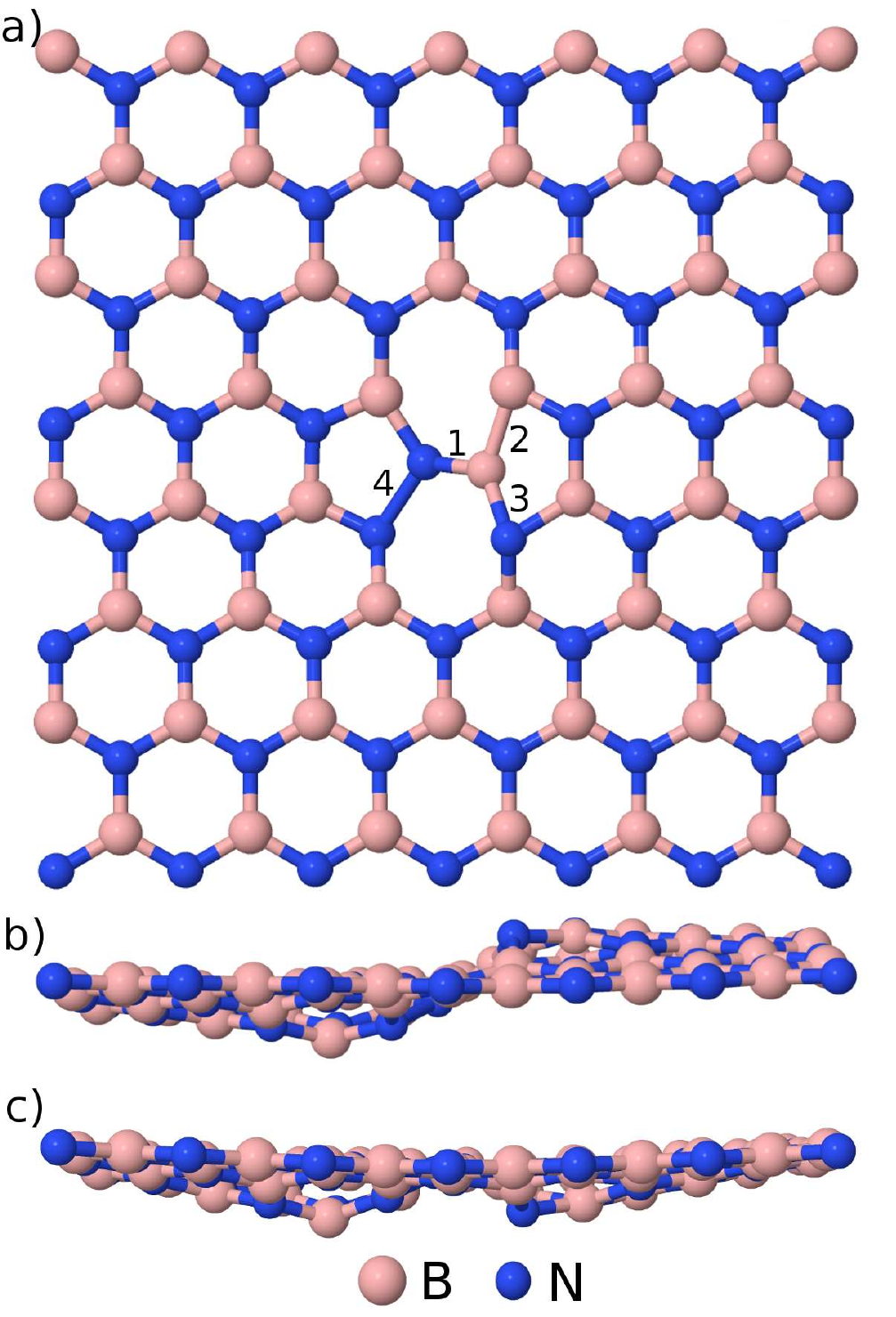}
\end{center}
\caption{Stone-Wales defects in BN. A pair of atoms is rotated by 90 degrees to form two pentagons and two heptagons (a). Side view of the two buckled structures with similar formation energy (see text): b) $SW_1$ . The bond lengths are 1) $1.47$ \r{A}, 2) $1.71$ \r{A}, 3) $1.75$ \r{A} and 4) $1.58$ \r{A}. c) $SW_2$.  The bond lengths 1 and 3 change slightly with respect to $SW1$: 1) $1.45$ \r{A} and 3 $1.72$ \r{A}. }
\label{fig:sw_def}
\end{figure}

First of all we consider Stone-Wales (SW) defects\cite{stone1986theoretical} where a 90 degrees rotation of a pair of atoms transforms four hexagons into two pentagons and two heptagons (see \fref{fig:sw_def}a)). In graphene SW defects  are known to have the lowest formation energy $\sim$ 4.7 eV \cite{carlsson2006structural}. Although SW defects have been theoretically studied in both nanotubes and single layer \textit{h}-BN\cite{ma2009stone,chen2009electronic}, they have not been found experimentally\cite{jin2009fabrication}. Also in our melting simulations we have not observed their formation. Recently, it has been suggested\cite{ma2009stone} that SW defects in graphene are further stabilized by a sine-like or cosine-like buckle and that this finding should hold also for other hexagonal single-layer crystals. 
By starting with a flat layer with a SW defect, and raising the temperature to 10 K  we find a spontaneous buckling to one of the two structures  $SW_1$  and $SW_2$ shown in \fref{fig:sw_def}b) and \fref{fig:sw_def}c) respectively which resemble the sine-like buckle proposed in \cite{ma2009stone}. We found the cosine-like buckle to be unstable and deform to the structure $SW_2$.
The formation energies (table \ref{table:for_energy_def}) of $SW_1$ is slightly lower and both are almost twice the value of a SW in graphene\cite{ma2009stone,carlsson2006structural}.

\begin{figure}[t]
\begin{center}
\includegraphics[angle=0,width=0.99\textwidth]{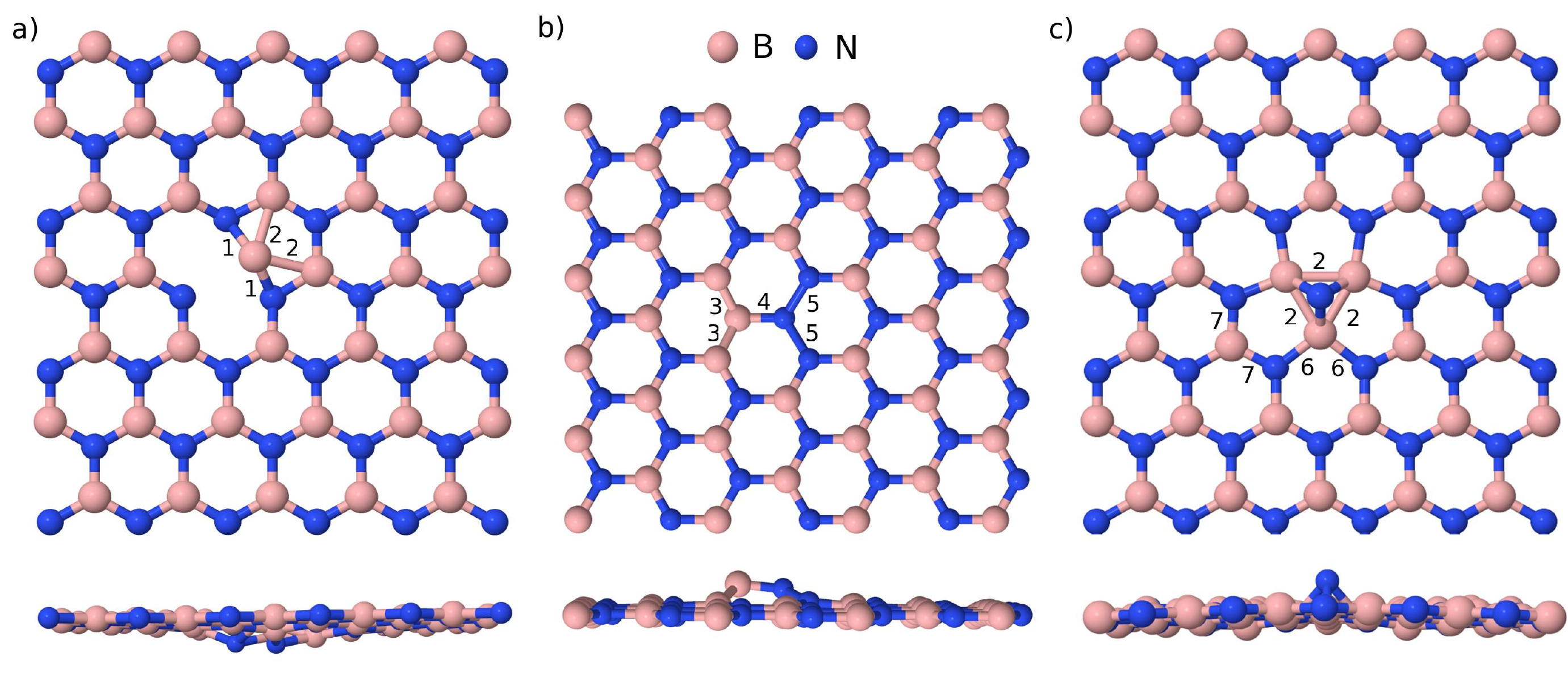}
\end{center}
\caption{top and side view of a) the BB defect, b) the antisite defect and c) the tetrahedron defect. Selected values of the interatomic distances: 1) 1.53 \r{A} 2) 1.90 \r{A} 3) 1.60 \r{A} 4) 1.47 \r{A} 5) 1.59 \r{A} 6) 1.61 \r{A} 7) 1.41 \r{A}.  In all three cases the defects only cause a local distortion in the z-direction leaving the rest of the sample relatively flat (see table~\ref{table:for_energy_def}).}
\label{fig:def_big}
\end{figure}

\begin{figure}[h!]
\begin{center}
\includegraphics[angle=0,width=0.5 \textwidth]{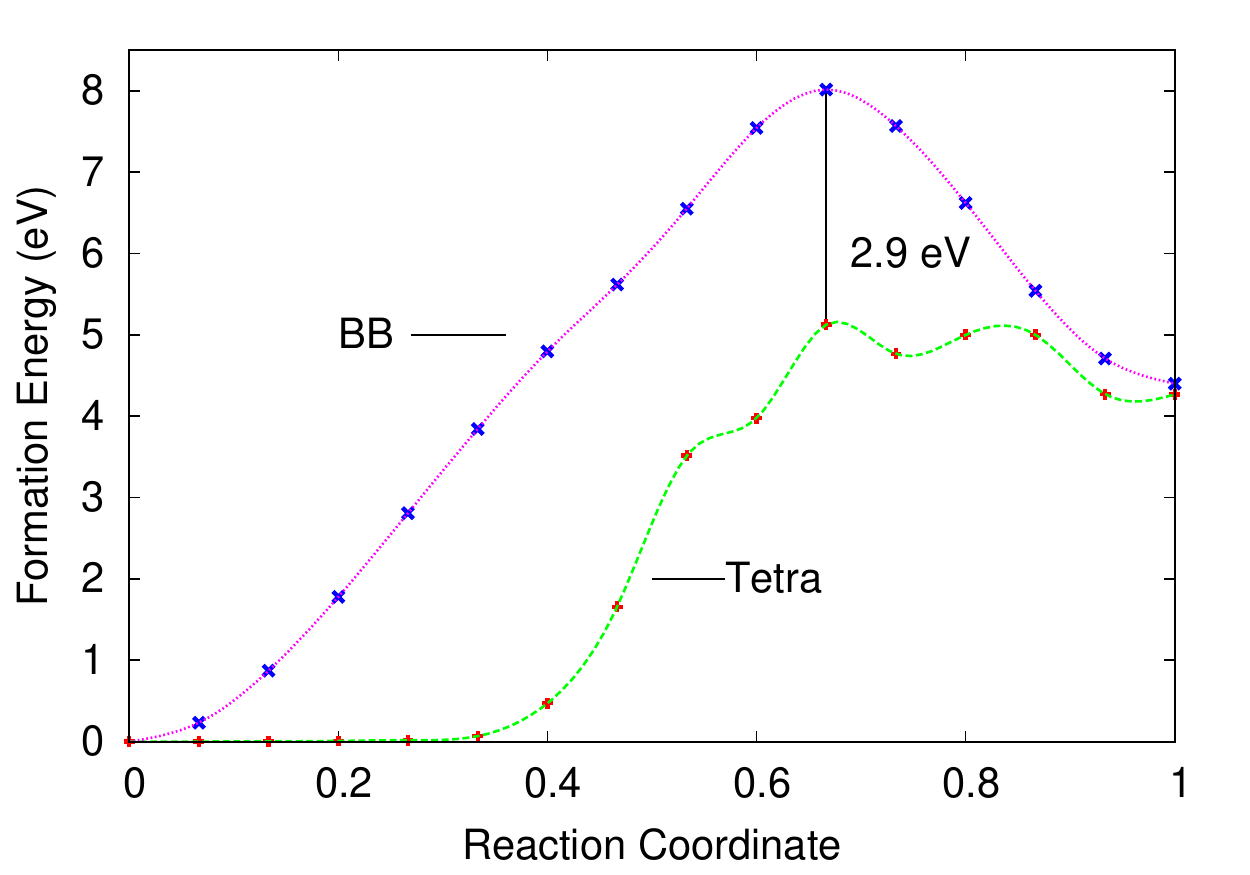}
\end{center}
\caption{The energy profile calculated using the nudged elastic band method for the BB and the tetrahedron defect. The reaction coordinate is only well defined at the points 0 where it represents the perfect sample and 1 for the final sample with the defect. The formation energy is defined to be $E_F = E_{defect} - E_{perfect}$. The energy difference between the two highest points is $2.9$ eV. }  
\label{fig:def_neb}
\end{figure}

While raising the temperature close to melting, we observe first two defects that have no counterpart in graphene, and turn out to have the lowest formation energy.
In \fref{fig:def_big}a) we show the defect that we call BB-defect because it results from a broken BN bond. 
The B atom moves to a metastable state with a slightly longer BN bond length (1.53 \r{A} instead of 1.47 \r{A}) with the two neighbours and the formation of two loose B-B bonds with a large bond length (1.9 \r{A}). No extended deviations in the z-direction occur for this defect (see table \ref{table:for_energy_def}). The formation energy ($4.4$ eV) is only $0.1$ eV higher than the tetrahedron defect (\fref{fig:def_big}c)) where a N atom pops up and form a NB$_3$ tetrahedron with the three B atoms in the plane. Also this defect does not lead to out of plane deformations in the surrounding. It has been suggested that a BN$_3$ tetrahedron structure with one N atom on top could explain some features in core level spectroscopy\cite{Peter2009,Pavlychev1998}. We find
that this structure is unstable and that the B atom abandons the layer if lifted from the plane. 
In \fref{fig:def_neb} we show the energy profile for the BB and tetrahedron defects calculated by the nudged elastic band (NEB) method, as implemented in \textsc{lammps}\cite{lammps}. We see that the barrier for the BB defect is $2.9$ eV higher than that of the  tetrahedron defect. In \fref{fig:def_big}b) we show also the antisite defect which is commonly found in semiconductors. As shown in table \ref{table:for_energy_def} its formation energy is intermediate between those of the BB and tetrahedron defects and those of the SW defects.

In table \ref{table:for_energy_def} we also give the maximum out of plane displacement and the average value of out of plane displacements in the whole sample. We can see that, with the exception of the antisite defect, the formation energy seems to be related to the out of plane distortions, as suggested for grain boundaries in graphene\cite{carlsson2011}. 

Lastly, we consider the various vacancies occurring when one or more atoms are removed. We define the vacancy giving as subscript the removed atoms, e.g. $V_N$ as a system missing one N atom. Apart from the $V_{BN}$ vacancy that keeps the stoichiometry, for these cases the evaluation of the formation energy requires to know the chemical potential of bulk phases of the constituents\cite{Chadi1988}. Different formation energies can result from use of different reference systems. Moreover, the chemical potential is often approximated by the cohesive energy, i.e. its value at $T=0$\cite{Chadi1988}. 
For BN, the formation energy of the $V_{BN}$, $V_{3B+N}$ and $V_{B+3N}$ vacancies has been calculated ab-initio in \cite{okada2009} by use of the chemical potentials of bulk metallic boron and solid nitrogen.
As discussed in Ref.\cite{Tersoff89}, this approach is not reliable for a phenomenological potential like the one we are using.
Therefore in table \ref{table:for_energy_vac} we just give the energy difference $\Delta E$ between the energy of the perfect sample ($E_{perfect}=n E_{coh}$) and the one of the sample relaxed after creation of the vacancy. 
We notice that $V_B$ and $V_N$ have the same $\Delta E$  because they both involve three BN broken bonds. Once the vacancy is created, no new bonds are formed.  The three atoms surrounding the vacancy remain two-fold coordinated with the same bond angle. The bond length with the  two nearest neighbours contracts from 1.46~\r{A} to 1.44~\r{A}.

Since, the single vacancies $V_N$ and $V_B$ have the same $\Delta E$, we have a qualitative indication of the formation energy by subtracting from $\Delta E$ the cohesive energy for each removed atom, irrespective of its nature, namely $A \equiv \Delta E-n_v E_{coh}$, where $n_v$ is the number of missing atoms. 

\begin{figure}[t]
\begin{center}
\includegraphics[angle=0,width=0.99\textwidth]{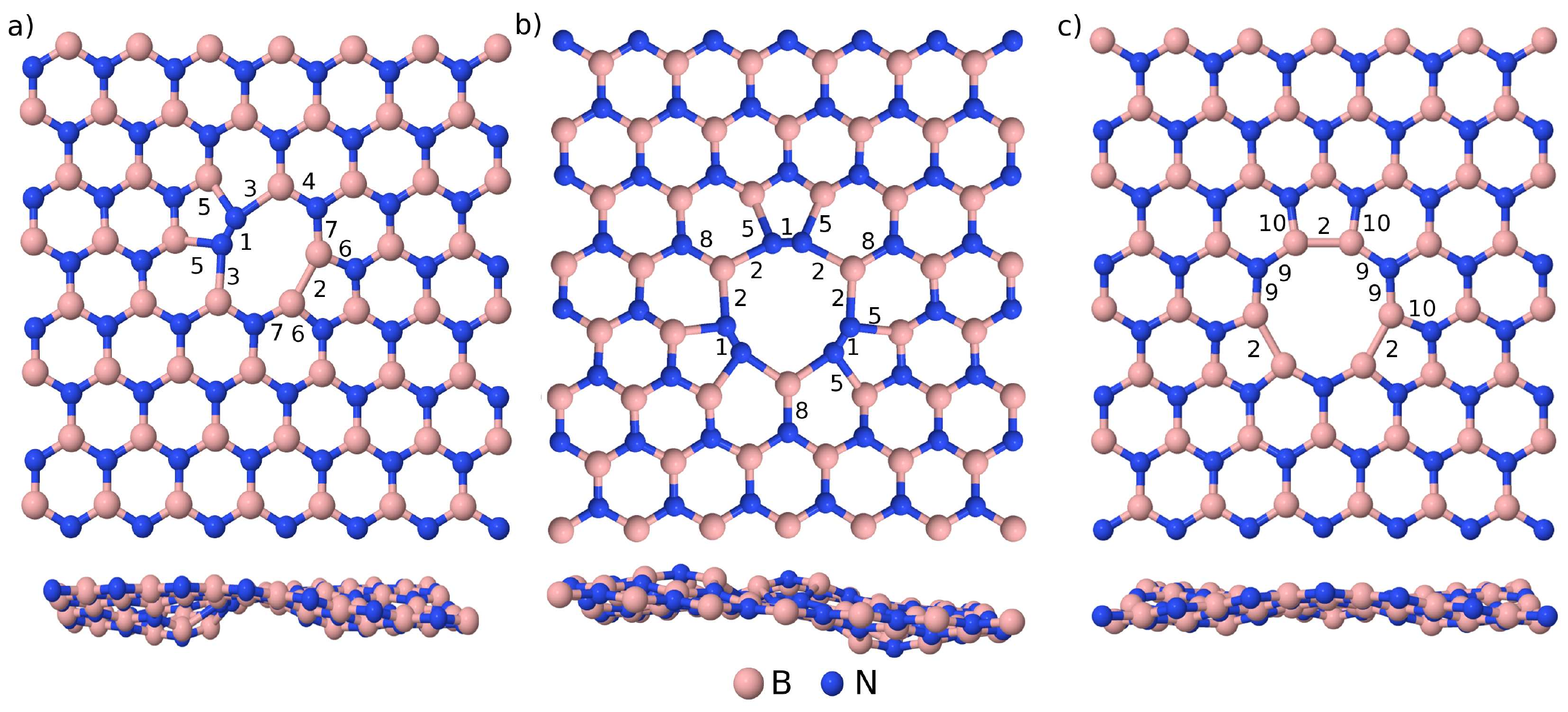}
\end{center}
\caption{ Top and side view of a) $V_{BN}$  b) $V_{3B+N}$ c) $V_{B+3N}$ where rebonding and ring formation is obtained by construction (see text and table\ref{table:for_energy_vac}). In both a) and b) short N-N bonds are created leading to strong out of plane distortions.  Selected values of interatomic distances: 1) 1.01 \r{A} 2) 1.91 \r{A} 3) 1.92 \r{A} 4) 1.47 \r{A} 5) 1.90 \r{A} 6) 1.49 \r{A} 7) 1.53 \r{A} 8) 1.58 \r{A} 9) 1.55 \r{A} 10) 1.52 \r{A}.}
\label{fig:vac_big}
\end{figure}

Also for  $V_{BN}$, $V_{3B+N}$ and $V_{B+3N}$ we find that no new bonds are formed after creation of the vacancy and that the structural changes are negligible.
In table~\ref{table:for_energy_vac} we give the corresponding values of $\Delta E$ and $A$. To establish whether these are indeed the lowest energy structures, we have brought the atoms around the vacancy closer to each other inducing the formation of new bonds re-establishing three-fold coordination as shown in \fref{fig:vac_big} for $V_{BN}$, $V_{3B+N}$ and $V_{B+3N}$. All these structures remain bonded after relaxation but have higher energy than the ones with no rebonding. This is due to the strong N-N bonds of only $\approx 1.01$ \r{A} in $V_{BN}$ and $V_{3B+N}$ that cause strong local out of plane distortions and hinder the ring formation. Indeed, the ring structure of the $V_{3B+N}$ vacancy shown in \fref{fig:vac_big}b) is obtained by construction and does not occur spontaneously whereas the $V_{B+3N}$, with no N-N bonds, may form spontaneously at finite temperature since it has an energy only marginally higher than the one without rebonding.

\section{Summary and conclusions}
\label{conclusion}
In summary, we have presented an extended study of the structural properties of single layer \textit{h}-BN by means of MD simulations based on the interatomic potential developed by Albe, M\"oller and Heinig and compared our findings to those for graphene. Validation of the results given by this potential opens the way to the development of a reliable potential capable to deal with hybrid BN-graphene structures. We find that the non-monotonic behaviour of the lattice parameter, the expansion of the interatomic distance and the growth of the bending rigidity with temperature are qualitatively similar to those of graphene. Conversely, the energetics of point defects is extremely different:  Stone Wales defects have formation energy twice as large as the two lowest energy defects in \textit{h}-BN which involve either a broken bond or an out of plane displacement of a N atom to form a tetrahedron with three B atoms in the plane. We have also studied the antisite defect and vacancies formed by one, two and four atoms. We hope that our results will stimulate further research on this topic.

\ack
This work is supported by FOM-NWO, the Netherlands. We thank Mehdy Neek-Amal, Merel van Wijk and Kostya Zakharchenko for discussions and Misha Katsnelson for his interest in this work.

\section*{References}


\clearpage 
\begin{table}[h!]
\caption{\label{table:for_energy_def} Formation energy $E_F= E_{defect} - E_{perfect}$ of defects that do not change the number of atoms. The height difference $h$ along the z-axis between the highest and lowest points is also given. In brackets we give the standard deviation  $[ \left<z^2\right>-\left<z\right>^2 ]^{1/2}$ evaluated over the whole sample.}
\begin{indented}
\item[]\begin{tabular}{@{}l l l}
\br
Defect& $E_F$ (eV) & $h$ (\r{A})\\
\mr
$SW_{1}$	&8.6 &2.26 (0.22)\\
$SW_{2}$	&8.8 &2.04 (0.28)\\
$BB$		&4.4 &1.12 (0.13)\\
Tetra	&4.3 &1.43 (0.20)\\
Anti-site &6.3 &1.41 (0.10)\\
\br
\end{tabular}
\end{indented}
\end{table}

\begin{table}[h!]

\caption{\label{table:for_energy_vac} The number $n_v$ of atoms removed to create the vacancy and the energy difference $\Delta E$ between the perfect sample and the sample with a defect for various vacancies. The quantity $A=\Delta E - n_v E_{coh}$ is given for the minimal energy structure without rebonding (see text), while $A'$ is the same quantity for the defects after ring formation (see figure \ref{fig:vac_big}). For the latter defects we also give the height difference $h$ along the z-axis between the highest and lowest points and in brackets the standard deviation $[ \left<z^2\right>-\left<z\right>^2 ]^{1/2}$ evaluated over the whole sample. }
\begin{indented}
\item[]\begin{tabular}{@{} l l l l l l}
\br
Defect&$n_v$ &$\Delta E$ (eV)&$A$ (eV) &$A'$ (eV) &$h$ (\r{A})\\
\mr
$V_{B}$		&1&11.7 &5.3 &\\
$V_{N}$		&1&11.7 &5.3 &\\
$V_{B+N}$	&2&19.7 &6.9 &8.8&2.06 (0.35)\\ 
$V_{B+3N}$	&4&36.0 &10.4&10.9&1.42 (0.31)\\ 
$V_{3B+N}$	&4&36.2 &10.6&16.1&2.63 (0.22)\\ 
\hline
\end{tabular}
\end{indented}
\end{table}

\end{document}